 \documentclass[smallabstract,smallcaptions]{dccpaper}

\usepackage{epsfig}
\usepackage{graphics}
\usepackage{amsmath}
\usepackage{amssymb}
\usepackage{microtype}
\usepackage{color}
\usepackage{url}
\usepackage[nolist]{acronym}
\usepackage{algorithm}
\usepackage{algpseudocode}
\usepackage{todonotes}
\usepackage{siunitx}

\newlength{\figurewidth}
\newlength{\smallfigurewidth}

\def\ve#1{\ensuremath{{\mathchoice{\mbox{\boldmath$\displaystyle #1$}}	%
			{\mbox{\boldmath$\textstyle #1$}}			%
			{\mbox{\boldmath$\scriptstyle #1$}}			%
			{\mbox{\boldmath$\scriptscriptstyle #1$}}}}}	%
\def\ma#1{\ensuremath{{\mathchoice{\mbox{\boldmath$\displaystyle #1$}}		%
			{\mbox{\boldmath$\textstyle #1$}}				%
			{\mbox{\boldmath$\scriptstyle #1$}}			%
			{\mbox{\boldmath$\scriptscriptstyle #1$}}}}}
\newcommand{\argmin}{\mathop{\mathrm{argmin}}}	
\newcommand{\norm}[1]{\ensuremath{ \Vert #1 \Vert}}
\newcommand{\RR}{\ensuremath{\mathbb{R}}}

\newcommand{\ZZ}{\ensuremath{\mathbb{Z}}}

\begin{document}
	
\begin{acronym}
	\acro{sqnr}[SQNR]{signal-to-quantization-noise-ratio}
	\acro{mpa}[DMP]{discrete matching pursuit}
	\acro{mcm}[MCM]{multiple constant multiplication}
	\acro{fpga}[FPGA]{field programmable gate array}
	\acro{dft}[DFT]{discrete fourier transform}
	\acro{lcc}[LCC]{linear computation coding}
	\acro{csd}[CSD]{canonically signed digit}
	\acro{lut}[LUT]{Lookup-Table}
\end{acronym}

\title
{\large
\textbf{Linear Computation Coding: \\ Exponential Search and Reduced-State Algorithms}
}

\author{%
Hans Rosenberger, Johanna S. Fröhlich, Ali Bereyhi and Ralf R. Müller\\[0.5em]
{\small\begin{minipage}{\linewidth}\begin{center}
\begin{tabular}{c}
	Institute for Digital Communications (IDC) \\
	Friedrich-Alexander-Universität Erlangen-Nürnberg \\
	Erlangen, Germany \\
	\makeatletter
	\{hans.rosenberger, johanna.froehlich, ali.bereyhi, ralf.r.mueller\}@fau.de
	\makeatother
\end{tabular}
\thanks{This work was supported by Deutsche Forschungsgemeinschaft (DFG) under the project Computation Coding (MU-3735/8-1).}
\end{center}\end{minipage}}
}

\maketitle
\thispagestyle{empty}

\vspace{5mm}

\begin{abstract}
Linear computation coding is concerned with the compression of multidimensional linear functions, i.e. with reducing the computational effort of multiplying an arbitrary vector to an arbitrary, but known, constant matrix. 
This paper advances over the state-of-the art, that is based on a \ac*{mpa} algorithm, by a step-wise optimal search.
Offering significant performance gains over \acs*{mpa}, it is however computationally infeasible for large matrices and high accuracy.
Therefore, a reduced-state algorithm is introduced that offers performance superior to \acs*{mpa}, while still being computationally feasible even for large matrices. 
Depending on the matrix size, the performance gain over \acs*{mpa} is on the order of at least \SI{10}{\percent}.
\end{abstract}

\Section{Introduction}
Multiplying a vector by a constant matrix is an ubiquitous task performed in various technical and scientific applications.
The main body of earlier work is focused on speeding up the calculation of matrix-vector multiplications in a structure-oriented fashion. 
A well-known example is the fast implementation of the \ac{dft}.
Here, the structure of the \ac{dft} matrix is exploited to eliminate redundant computations and reduce the number of required operations as compared to a naive implementation.
For arbitrary constant matrices, redundancies within the finite-precision representation of the matrix entries can be exploited as well, a method that is typically known as common subexpression sharing/elimination.
Earlier work in this respect has either targeted special cases of constant multiplication~\cite{Thong2011,Voronenko2007} or has proposed schemes with high computational complexity, such that their implementation in practice is difficult for medium to large size matrices~\cite{Boullis2003,Aksoy2016}.

Recently, \ac{lcc} has been proposed in~\cite{mueller2020sparse,mueller2021lcc,muellerlcc2022}.
This framework develops an information-theoretic scheme for the efficient calculation of matrix-vector products that is especially well-suited for the implementation on reconfigurable hardware, such as \acp{fpga}~\cite{Lehnert2022}.
Similar to rate-distortion theory, \ac{lcc} is concerned with the tradeoff between distortion and compression.
However, instead of compressing data, \ac{lcc} deals with the lossy compression of multidimensional linear functions under a given fidelity constraint.
An instance can be found in~\cite{muellerlcc2022}, where an optimal decomposition scheme is first defined in terms of classical metrics for computation and distortion.
A greedy approach is then developed to approximate the proposed scheme sub-optimally with tractable complexity. 

\SubSection{Contributions}
In this paper, we develop a new \ac{lcc} scheme.
Similar to earlier approaches discussed in~\cite{muellerlcc2022}, the optimal decomposition deals with an exponentially complex problem.
We first address this problem via an exhaustive search procedure with a careful optimization.
This enables us to evaluate the performance of the optimal scheme for reasonable matrix sizes.
We then present a computationally tractable scheme by proposing a reduced-state algorithm for the underlying search problem.
Our investigations show that the proposed algorithm can achieve a computation-distortion tradeoff close to the exponentially-complex optimal scheme while drastically reducing the decomposition complexity.

\SubSection{Notation}
Vectors are denoted as lower-case boldface letters $\ve{x}$ and matrices as upper-case boldface letters $\ma{X}$.
The Euclidean and the Frobenius norm are denoted by $\norm{\cdot}_2$ and  $\norm{\cdot}_\mathrm{F}$, respectively.
The symbol $\ve{0}_{N \times K}$ denotes an $N \times K$ matrix with all zero elements, $\ma{I}_{N \times K}$ denotes the augmented identity matrix of dimension $N \times K$ and $\ve{1}_{j,K}$ denotes the $j$-th row unit vector in $K$ dimensions.


\Section{Problem Formulation}

We consider the problem of matrix-vector multiplication, i.e the calculation
\begin{align}\label{eq::matprod}
	\ve{y} = \ma{A}\ve{x}
\end{align}
for an arbitrary input vector $\ve{x} \in \RR^{K\times 1}$ and a constant matrix $\ma{A} \in \RR^{N\times K}$. 
Commonly, matrices are approximated by quantizing their entries independently. 
By using the \ac{csd} binary representation the quantization error can be decreased on average by a factor of $\sqrt{28}$ per \ac{csd}~\cite{Booth1951}.
This still leaves room for improvement. 
\ac{lcc} instead suggests to approximate $\ma{A}$ by a product of matrices, i.e. finding $\ma{W}$ and $\ma{C}$ such that
\begin{align}
	\ma{A} \approx \ma{W} \ma{C}.
\end{align}
The matrix $\ma{C} \in \mathcal{A}^{N \times K}$ is termed the codebook matrix and $\ma{W} \in \mathcal{A}^{N \times N}$ is termed the wiring matrix in the sequel\footnote{In~\cite{mueller2021lcc} the multiplication order of the decomposed matrices is reversed. Please note that this change makes no difference to the general idea of the decomposition and to the following algorithms. It is equal to the transposed version of the algorithm presented in~\cite{muellerlcc2022}.}.
The entries of the wiring matrix are restricted to the set of zero and signed powers of two ($\mathcal{A} \subseteq \{ 0, \pm 2^\ZZ \}$).

Obtaining the wiring and codebook matrix jointly is typically NP-hard and infeasible.  
To overcome this computational intractability,~\cite{muellerlcc2022} proposes a scheme where the $n$-th row of the wiring matrix is determined by solving the following sparse recovery problem for some design parameter $S<N$ controlling the cost between distortion and computation effort~\cite{Foucart2013}
\begin{align}\label{eq::columnopt}
	\ve{w}_{n} = \underset{\ve{\omega} \in \{ \ve{\omega} = \sum_{s=1}^S i_s \ve{1}_{j_s,N}: \; i_s \in \mathcal{A},\, j_s \in \{ 1,...,N \} \; \forall s \} }{\argmin}
	\norm{\ve{a}_n - \ve{\omega} \ma{C}}_2  \quad \forall n.
\end{align}
The new scheme is still NP-hard, but not in $N$, anymore, but in $S$.
Thus, small values of $S$ are required, in practice.

In order to have a high accuracy despite small values of $S$, the factorization procedure can be applied multiple times.
Then the product $\ma{C}_i = \ma{W}_i\ma{C}_{i-1}$ of the previous wiring step acts as the new codebook for obtaining the following matrix factor $\ma{W}_i$ of the current wiring step.
Hence, by setting\footnote{In~\cite{mueller2021lcc} this choice is termed the self-designing codebook. It was found to work very well for a wide range of matrices.} $\ma{C}_0 = \ma{I}_{N \times K}$, we obtain the approximated matrix $\ma{P}$ after $I$ wiring steps:
\begin{align}\label{eq::seqmat}
	\ma{A} \approx \ma{P} = \left( \prod_{i=1}^{I} \ma{W}_i \right) \ma{C}_0.
\end{align}

To quantify the accuracy of a given approximation $\ma{P}$ we use the \ac{sqnr}
\begin{align}\label{eq::sqnr}
\mathrm{SQNR} (\ma{A}, \ma{P}) = \frac{\norm{\ma{A}}_\mathrm{F}^2}{\norm{\ma{A}-\ma{P}}_\mathrm{F}^2}.
\end{align}

\SubSection{Computational Cost}
In a binary number representation the multiplication by a signed power of two corresponds only to a bitshift.
On reconfigurable hardware, this shift can be realized simply by appropriate wiring without the need for dedicated processing elements such as adders~\cite{Lehnert2022}.
The parameter $S$ in~\eqref{eq::columnopt} determines the number of vectors from the codebook to be used in forming the linear combination to approximate a row $\ve{a}_n$ of $\ma{A}$.
It therefore directly controls the computational cost, as in computing the linear combination, exactly $S-1$ additions are required.
No multiplications, except by signed powers of two, are necessary due to the specific structure of the wiring matrix.
Therefore, the separate product of the decomposed matrices with the input vector $\ve{y} \approx \ma{W} (\ma{C} \ve{x})$ is much simpler to compute than calculating the product in \eqref{eq::matprod} straightforwardly.

The total computational cost $C_\mathrm{add}$ of a decomposition in~\eqref{eq::seqmat} is given by the number of additions (or subtractions) required to form the linear combinations
\begin{align}
	C_\mathrm{add} = I N (S-1).
\end{align}
\Section{Algorithms}

In this section we will briefly look at the state of the art for solving the optimization problem in~\eqref{eq::columnopt} to obtain the wiring matrices and then introduce two improved novel algorithms.

%

\SubSection{State-of-the-Art: Discrete Matching Pursuit}
The \ac{mpa} follows the matching pursuit approach to successively determine the wiring coefficients. The algorithm can be summarized in the following key steps; for details see~\cite{mueller2021lcc}. 
\begin{enumerate}
	\item Start with iteration $s \gets 0$. Initialize $\ve{\omega} \gets \ve{0}_{1 \times N}$ 
	\item Update $\ve{\omega}$ in at most a single component, such that $\norm{\ve{a_n} - \ve{\omega}\ma{C}}_2$ is minimized.
	\item Increment $s$.
	\item If $s \leq S$, go to step 2, otherwise the procedure terminates.
\end{enumerate}
It is straightforward to show that the time complexity of the \ac{mpa} algorithm for computing a single matrix factor scales with $\mathcal{O}(N^3 S)$.

%
%


\SubSection{Exponential Search Algorithm}
The row-wise optimization problem in~\eqref{eq::columnopt} is NP-hard. 
However, for small $S$, reasonable matrix sizes and some careful optimization it can be solved in a tractable timeframe.
We limit the set of scaling factors to a finite set of signed powers of two ($\mathcal{A}_\mathrm{exp} \subset \{ 0, \pm 2^\ZZ\}$), as an exhaustive search over the whole set is infeasible.
As the search procedure has to be performed for each row of the target matrix individually, the time complexity for the computation of each wiring step is given by $\mathcal{O}(N^{S}|\mathcal{A_\mathrm{exp}}|^{S})$.

Generally, which and how many coefficients are included in the subset $\mathcal{A}_\mathrm{exp} $ is a design parameter and needs to be adapted to each specific decomposition.
It depends primarily on two factors. 
First, the current wiring step plays a crucial role. 
For each additional wiring layer, the error between each row of the target matrix and the approximated matrix decreases.
Hence, for any subsequent wiring step, smaller coefficients are needed to scale the rows of the newly found codebook matrix to appropriately approximate the residual error.
This also means that for high desired accuracy the coefficient set needs to be chosen large, i.e. to include also many small coefficients, to accurately approximate the error.
Furthermore, relative variations in the length of the row vectors of the target matrix require a larger coefficient set to compensate for differences.
Still, to keep the decomposition computationally feasible, the number of elements in $\mathcal{A}_\mathrm{exp} $ needs to be chosen as small as possible, as the computational complexity scales exponentially in $S$ with the product of the size of the coefficient set $|\mathcal{A}_\mathrm{exp}|$ and $N$ as base. 

A promising approach for further research is to adapt the coefficient set for each wiring step dynamically based on the current fidelity of the approximation.
The coefficient set may then be determined from the probability distribution of the likely entries of the wiring matrix.


\SubSection{Reduced-State Algorithm}
The runtime of the exponential algorithm for a matrix with $N=64$ rows is on the order of an hour\footnote{For a multithreaded implementation in Python with Numba acceleration executed on an Intel i9-12900@\SI{2.40}{\giga\hertz} and parameters $S=3$, $\mathcal{A}_\mathrm{exp} = \left\{ \pm 2^{-40}, \dots, \pm 2^3 \right\}$.}, for larger matrices the runtime scales up accordingly and can become infeasible.
Hence, for some applications the computational burden of the exponential algorithm might consequently not be feasible.
A reasonable compromise between complexity and performance is desirable.

Unlike in \ac{mpa}, we do not immediately update the components of the wiring vector for the reduced-state algorithm.
Instead, we keep updating in each iteration a list of the $M$ best vectors, which minimize~\eqref{eq::columnopt} and select at the termination of the algorithm the vector with minimum error from the list.
Specifically, we apply the following successive, greedy procedure for the optimization problem in \eqref{eq::columnopt}:
\begin{enumerate}
	\item Start with $s \gets 0$. Initialize the set $\Omega$ with $M$ all-zero vectors.
	\item For each $\ve{\omega} \in \Omega$ find the set $\Omega_m$ of $M$ mutually distinct vectors $\ve{\omega}_{\Tilde{m}}$ with $\Tilde{m} \in \{1,...,M\}$ that minimize $\norm{\ve{a}_n - \ve{\omega}_{\Tilde{m}}\ma{C}}_2$ and differ from $\ve{\omega}$ in at most a single component.
	\item Update the set $\Omega$ by selecting from $\bigcup_{m=1}^M \Omega_m$ the $M$ distinct vectors $\ve{\omega}$ with minimum $\norm{\ve{a}_n - \ve{\omega}\ma{C}}_2$
	\item Increment $s$.
	\item If $s \leq S$, go to step 2, else, continue
	\item Return $\underset{\ve{\omega} \in \Omega}{\argmin} \norm{\ve{a}_n - \ve{\omega} \ma{C}}_2$
\end{enumerate}
By tuning $M$, we are able to adjust the space of possible combinations that the algorithm explores.
For each wiring step, we have to evaluate $\mathcal{O}(SN^3M^2)$ combinations. 
Compared to \ac{mpa} the complexity is increased by a factor of $M^2$.
Note that for $M=1$ the algorithm reduces to the \ac{mpa}. 


%

\Section{Numerical Evaluation}
In this section, we compare the performance of the proposed algorithms to the baseline \ac{mpa}~\cite{muellerlcc2022}.
We decompose matrices whose entries are drawn i.i.d. from a Gaussian distribution with zero mean and unit variance. 
Similar to \ac{mpa}, the performance of the improved versions of the algorithm hardly depends on the distribution of the matrix elements.
The algorithms are invariant to scaling of the variance, however it is crucial that for the exponential search algorithm the coefficient set $\mathcal{A}_\mathrm{exp}$ is scaled appropriately as well, as to not compromise performance.
Throughout the simulations, we select the coefficient set for the exponential search algorithm to $\mathcal{A}_\mathrm{exp} = \{ \pm 2^{-40},..,\pm 2^3 \}$.

\begin{figure}[t]
	\centering
	\includegraphics[width=\textwidth]{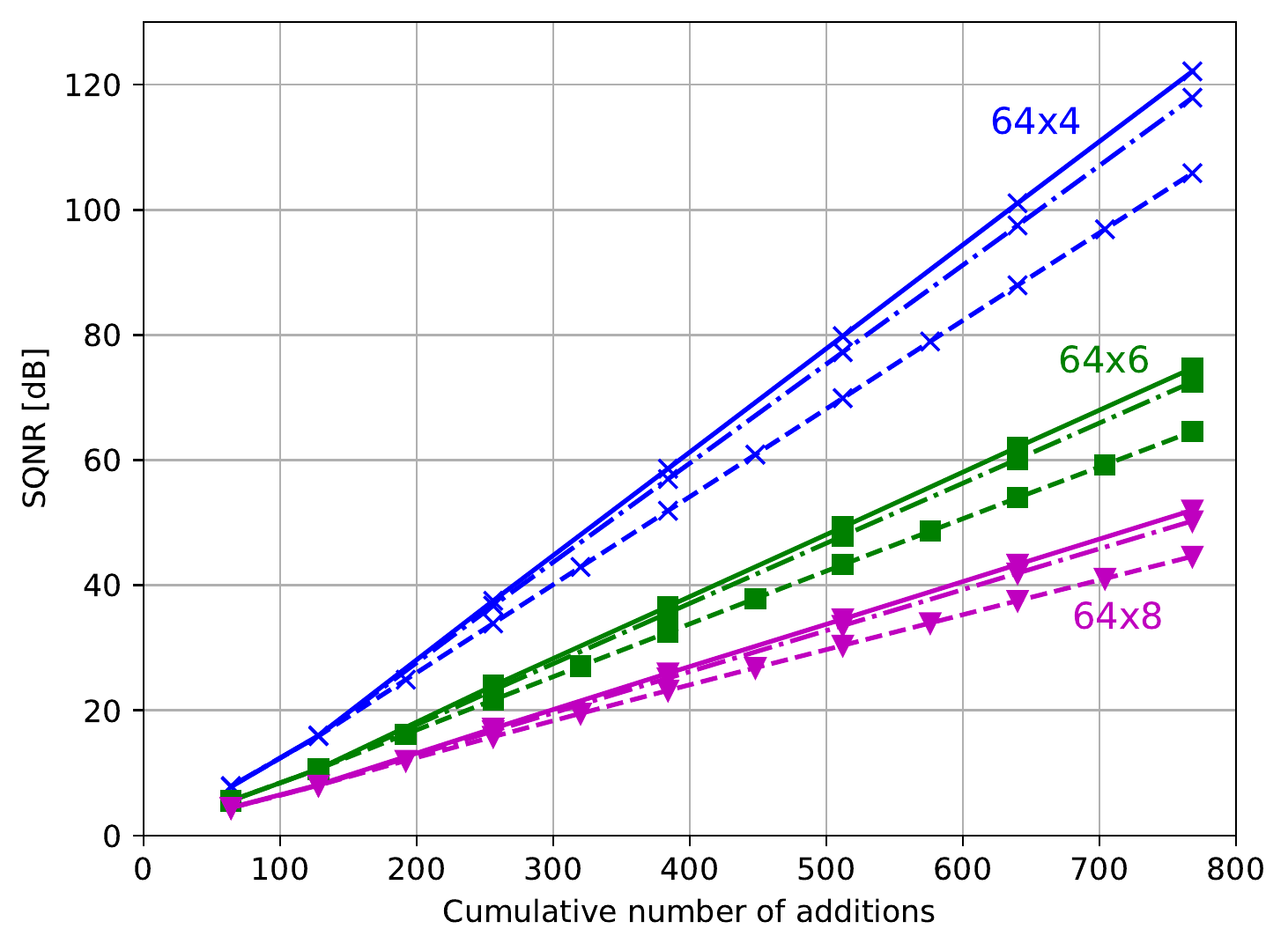}
	\caption{Performance comparison for matrices with 64 rows and different aspect ratios. Matrices of dimension $64 \times 4$ are indicated by crosses, $64 \times 6$ is indicated by squares and $64 \times 8$ is indicated by triangles. The solid lines refer to the exponential search algorithm ($S=3$), the dashed lines to the \ac{mpa}~\cite{muellerlcc2022} ($S=2$) and the dashed-dotted lines to the reduced state algorithm with memory size $M=10$ and $S=3$. Results are averaged over $10^4$ matrix entries for the exponential search algorithm and over $10^5$ matrix entries for the other algorithms.}
	\label{fig::ratecomp}
\end{figure}

As a first experiment, we compare all three algorithms for fixed matrix sizes in Figure~\ref{fig::ratecomp}.
We choose matrices with $N=64$ rows and vary the number of columns $K$ from four to eight.
Hence, we can compare the performance for different aspect ratios of the matrices\footnote{\ac{lcc} works best for matrices with an exponential aspect ratio, i.e. for $K \approx \log_2 N$. For square matrices it is beneficial to cut these into multiple tall matrices and decompose each slice independently, see~\cite{Lehnert2022} for details.}.
To quantify performance, we plot the tradeoff between distortion and computational cost. 
The latter being measured by the number of cumulative additions $C_\mathrm{add}$ required for a given wiring step.

As Figure~\ref{fig::ratecomp} shows, for all three matrix sizes there is a performance gain by both proposed algorithms against \ac{mpa}.
Further, for memory size $M=10$, the reduced state algorithm performs only slightly worse than exhaustive search.

For the first two wiring steps, the performance of all three algorithms is equal for a given matrix size.
This is due to the fact that for the first two steps we use \ac{mpa} with $S=2$ for an initial refinement of the codebook\footnote{Using any of the novel algorithms with $S=2$ is possible as well with very similar performance, i.e. a slight increase in \ac{sqnr} by \SI{0.2}{\decibel} to \SI{0.5}{\decibel} for the novel algorithms and matrix sizes considered.}.
Applying any of the algorithms with $S \geq 3$ directly to the initial codebook $\ma{C}_0 = \ma{I}_{N \times K}$ would lead to degraded performance for the first few wiring steps.
Instead it is beneficial to set $S=2$ to allow for more frequent updates of the codebook in the beginning.
From empirical investigations it seems that two wiring iterations with $S=2$ are most beneficial for the overall performance.

\begin{figure}[t]
	\centering
	\includegraphics[width=\textwidth]{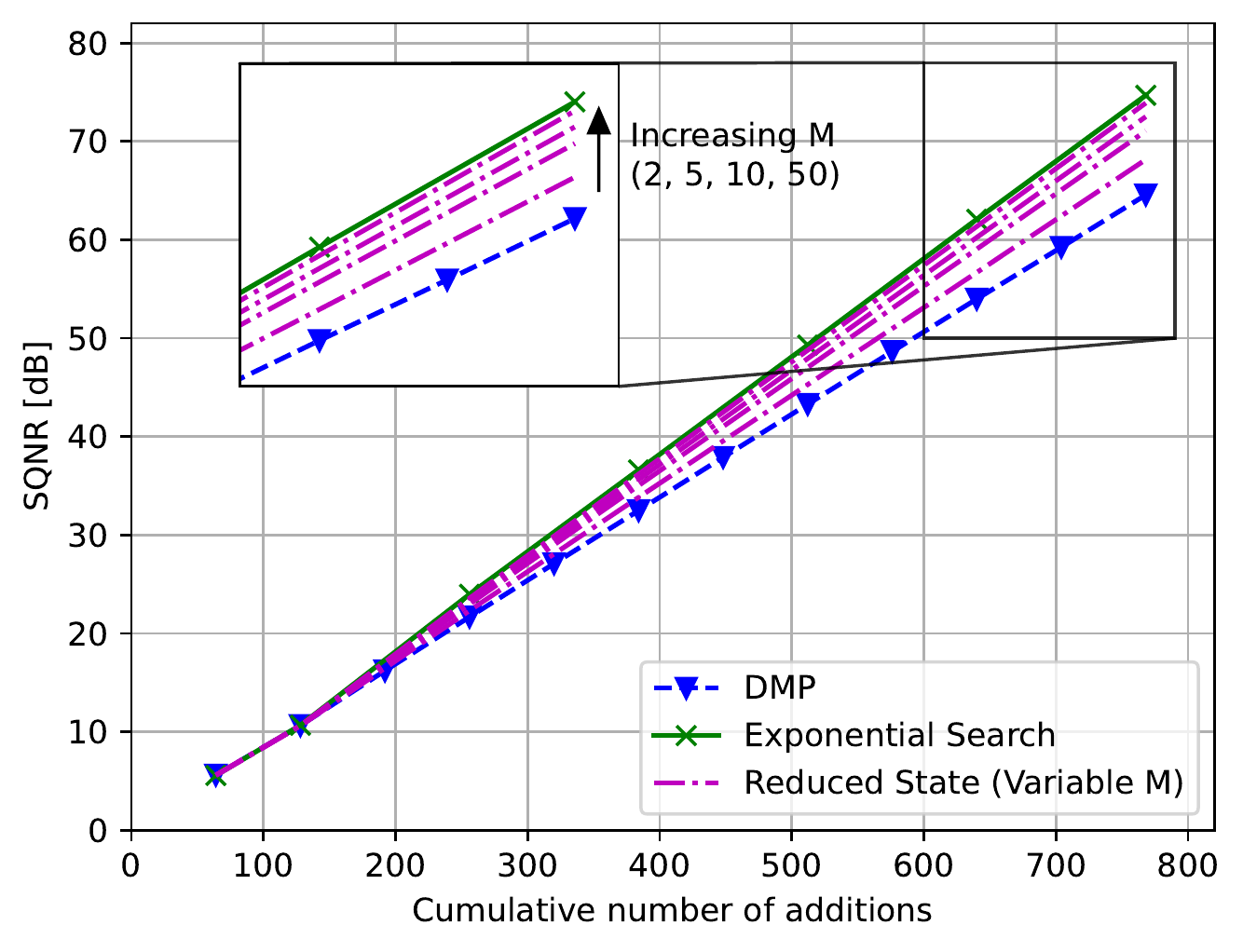}
	\caption{Performance comparison for different memory sizes $M$ of the reduced state algorithm of a $64 \times 6$ matrix. The solid line refers to the exponential search algorithm ($S=3$), the dashed line to the \ac{mpa}~\cite{muellerlcc2022} ($S=2$) and the dashed-dotted lines to the reduced state algorithm with 4 different memory sizes and $S=3$. Results are averaged over $10^4$ matrix entries for the exponential search algorithm and over $10^5$ matrix entries for the other algorithms.}
	\label{fig::memorycomp}
\end{figure}

As the next experiment, we compare the performance of the reduced state algorithm for different choices of the memory parameter $M$ for given matrix size of $64 \times 6$ in Figure~\ref{fig::memorycomp}.
From the figure we observe that even for small $M$ the reduced state algorithm offers a noticeable performance gain over \ac{mpa}.
As $M$ grows, the reduced state algorithm approaches the performance of the exponential search.
The large advantage of the reduced state algorithm is that the computation time is reduced drastically\footnote{For the considered matrix size the execution time differs from an hour for the exponential algorithm to a few seconds for the reduced state algorithm in practice.}.
For $M=1$ the reduced state algorithm reduces to \ac{mpa} and both lines coincide for any given matrix size.

\begin{figure}[t]
	\centering
	\includegraphics[width=\textwidth]{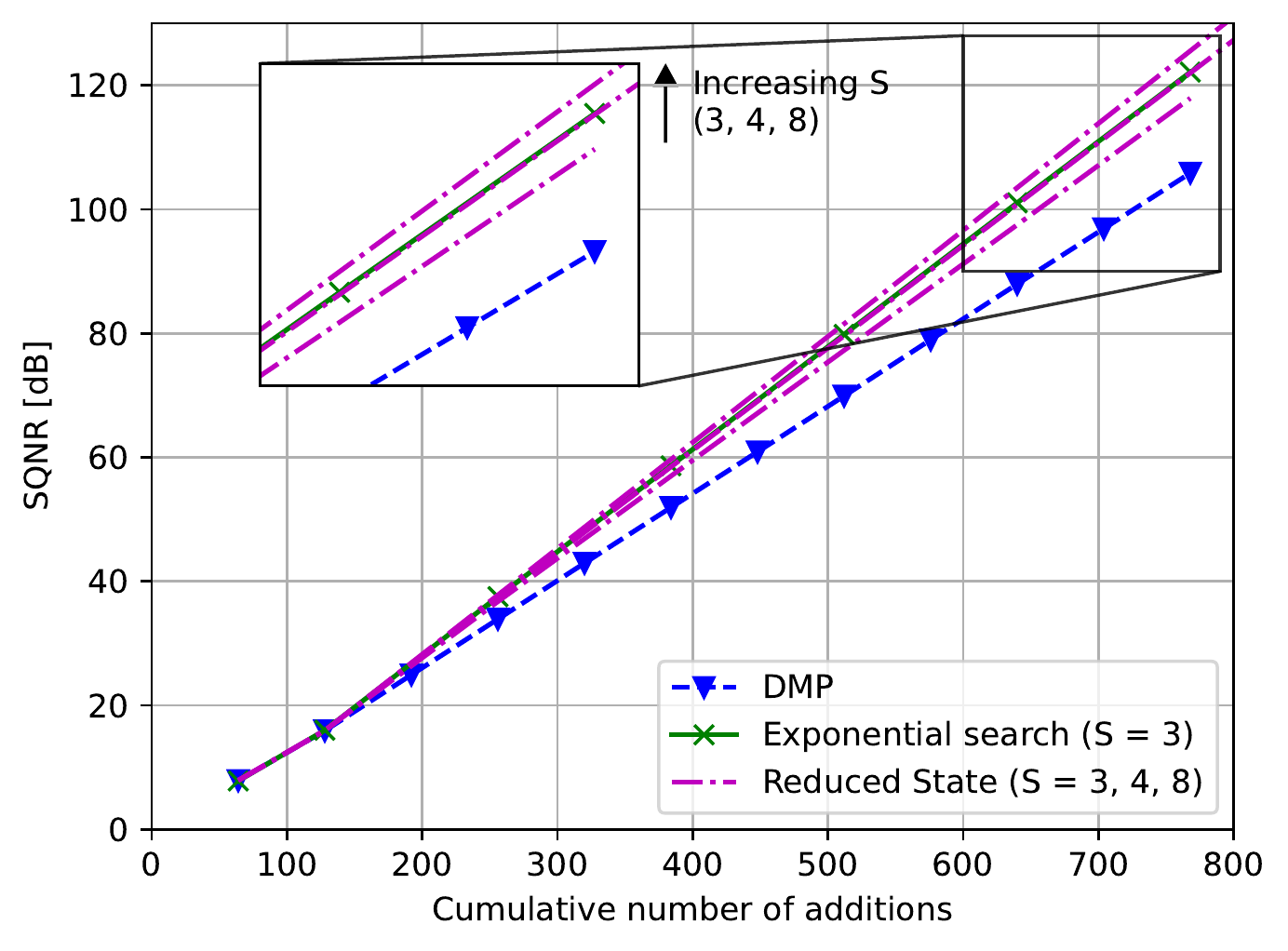}
	\caption{Performance comparison for different choices of the parameter $S$ of a matrix with dimension $64 \times 4$. The solid line refers to the exponential search algorithm ($S=3$), the dashed line to the \ac{mpa}~\cite{muellerlcc2022} ($S=2$) and the dashed-dotted lines to the reduced state algorithm with  different choices of the parameter $S$ and $M=10$. Results are averaged over $10^4$ matrix entries for the exponential search algorithm and over $10^5$ matrix entries for the other algorithms.}
	\label{fig::stepcomp}
\end{figure}


Table~\ref{tab::THETABLE} lists the relative performance gains over \ac{mpa} for various matrix sizes and configurations of the algorithms.

\begin{table}[tp]
	\centering
	\caption{\label{tab::THETABLE}Relative average gain in terms of \ac{sqnr} of the novel algorithms over the baseline \ac{mpa} algorithm with $S=2$ (For at least \SI{8}{bit} signed integer accuracy ($10\log(\textrm{SQNR}) \geq \SI{47}{\decibel}$)). Results are averaged over $10^4$ matrix entries for the exponential search algorithm and over $10^5$ matrix entries for the other algorithms.}
	\begin{tabular}{|c|c|c|c|c|c|c|c|}
		\hline
		& Exponential & \multicolumn{6}{c|}{Reduced State} \\
		\cline{3-8}
		& search & \multicolumn{2}{c|}{$S=3$} & \multicolumn{2}{c|}{$S=4$} & \multicolumn{2}{c|}{$S=8$} \\
		\cline{3-8}
		Matrix size & $S=3$ & $M=5$ & $M=10$ & $M=5$ & $M=10$ & $M=5$ & $M=10$ \\
		\hline
		$16 \times 2$ & \SI{17.8}{\percent} & \SI{10.4}{\percent} & \SI{13.4}{\percent} & \SI{14.1}{\percent} & \SI{17.8}{\percent} & \SI{16.7}{\percent} & \SI{21.9}{\percent} \\
		$16 \times 4$ & \SI{34.5}{\percent} & \SI{16.0}{\percent} & \SI{24.5}{\percent} & \SI{25.8}{\percent} & \SI{32.7}{\percent} & \SI{25.4}{\percent} & \SI{34.4}{\percent} \\
		\hline
		$32 \times 4$ & \SI{15.7}{\percent} & \SI{10.5}{\percent}  & \SI{12.9}{\percent} & \SI{14.0}{\percent} & \SI{17.2}{\percent} & \SI{18.4}{\percent} & \SI{22.3}{\percent} \\
		$32 \times 6$ & \SI{25.3}{\percent} & \SI{15.5}{\percent} & \SI{19.0}{\percent} & \SI{19.7}{\percent} & \SI{24.5}{\percent} & \SI{19.4}{\percent} & \SI{26.0}{\percent} \\
		\hline
		$64 \times 4$ & \SI{12.9}{\percent} & \SI{7.5}{\percent} & \SI{9.8}{\percent}& \SI{11.0}{\percent} & \SI{13.8}{\percent} & \SI{13.5}{\percent} & \SI{16.8}{\percent} \\
		$64 \times 6$ & \SI{14.9}{\percent} & \SI{9.6}{\percent} & \SI{11.4}{\percent} & \SI{13.6}{\percent} & \SI{16.5}{\percent} & \SI{15.6}{\percent}  & \SI{19.0}{\percent}  \\
		\hline
	\end{tabular}
\end{table}

\SubSection{Practical Considerations for the Choice of $S$}
Figure~\ref{fig::stepcomp} shows the performance of the reduced state algorithm for varying $S$.
Due to the exponentially growing complexity in $S$, the exponential algorithm is not feasible for $S>3$, except for very small matrices.
We can observe, that by choosing $S=4$ for the reduced state algorithm, we approximately achieve the same distortion-cost tradeoff as for the exponential algorithm with $S=3$.
For choosing $S$ even larger the gains increase likewise.

However for a practicable implementation $S$ should not be chosen arbitrarily.
In~\cite{Lehnert2022}, the performance of \ac{mpa} is validated in an implementation on reconfigurable hardware with $S=2$.
This means that on an \ac{fpga} exactly $N$ adders are required per wiring matrix.
With the inputs depending only on the outputs of the previous wiring matrix, the decomposition is well suited for parallel execution and pipelining~\cite{Lehnert2022}.
Due to the greedy, step-wise nature of \ac{mpa} $S>2$ does not offer significant performance gains over $S=2$.
Both novel algorithms behave differently.

From an implementation point of view, $S=3$ is even more suitable than $S=2$ for an effective implementation in hardware due to the availability of efficient adders with three inputs~\cite{simkins2007ternadd}.
Interestingly, on modern \acp{fpga}, these adders do not require more hardware resources, in terms of \acp{lut}, than an adder with two inputs.
Any powers of two and three ($S=4,8,9,\dots$) can be realized efficiently as well by the use of adder trees.
However, it is questionable if choosing $S>4$ is beneficial, as performance gains over $S=3$ or $S=4$ are small and the desired fidelity of the approximation cannot be chosen in a fine granularity anymore\footnote{For a matrix of dimension $64\times4$ the reduced state algorithm with $S=8$ improves the \ac{sqnr} approximately by \SI{70}{\decibel} per matrix factor.}.

\Section{Conclusion}
In this paper, we have proposed two new algorithms for \ac{lcc}, a framework for the lossy compression of multidimensional linear functions.
While the exponential search algorithm shows the best performance, it is generally infeasible especially for large matrices.
The proposed reduced-state algorithm, performs close to exponential search at a fraction of the computational complexity.
The time complexity of the decomposition compared to the baseline algorithm from earlier works is only mildly increased, while the performance gains over the baseline \ac{mpa} algorithm are on the order of at least \SI{10}{\percent}. 

\Section{References}
\bibliographystyle{IEEEbib}
\bibliography{refs}

\end{document}